\documentclass[twoside]{article}
\usepackage{graphicx,amssymb,mathrsfs,amsmath}
\catcode`@=11
\long\def\@makefntext#1{\noindent #1}
\newskip\tabcentering \tabcentering=1000pt plus 1000pt minus 1000pt
\def\REF#1{\par\hangindent\parindent\indent\llap{#1\enspace}\ignorespaces} 
\def\MCH#1#2{\setbox0=\hbox{\raise#1\hbox{#2}}\smash{\box0}}

\def\@evenfoot{}\def\@oddfoot{}

\def\@evenhead{\hbox to\textwidth{\footnotesize\rm\thepage \hfill
{\it Yong Yao, Jia Xu, Lu Yang}}} 

\def\@oddhead{\hbox to \textwidth{\footnotesize{
} \hfill\thepage}}

\def\sec#1{\vskip 3mm\leftline{\bf #1}\vskip 1mm}

\catcode`@=12

\def\bc{\begin{center}}
\def\ec{\end{center}}
\def\no{\noindent}
\def\hang{\hangindent\parindent}
\def\textindent#1{\indent\llap{\qquad #1\ \ \enspace}\ignorespaces}
\def\ref{\par\hang\textindent}

\begin{document}
\abovedisplayskip=6pt plus 1pt minus 1pt \belowdisplayskip=6pt
plus 1pt minus 1pt
\thispagestyle{empty} \vspace*{-1.0truecm} \noindent
\vskip 10mm

\bc{\large\bf A Successive Resultant Projection\\
\vskip 2mm
 for Cylindrical Algebraic Decomposition

\footnotetext{\footnotesize }

\footnotetext{\footnotesize The
work  was partially supported by the National Key Basic Research Project of China (GrantNo. 2011CB302402)
and the Fundamental Research Funds for the Central Universities, Southwest University for
 Nationalities (GrantNo.12NZYTH04).}} \ec

\vskip 5mm
\bc{\bf Yong Yao \\
{\small\it Chengdu Institute of Computer Applications, Chinese
Academy of Sciences, Chengdu,\\ Sichuan 610041, PR China \quad
E-mail: yaoyong@casit.ac.cn}}\ec

\vskip 5mm
\bc{\bf Jia Xu\\
{\small\it College of Computer Science and Technology, Southwest
University for Nationalities, Chengdu, Sichuan 610041, PR China
\quad E-mail: jjiaxu@gmail.com}}\ec

\vskip 5mm
\bc{\bf  Lu Yang \\
{\small\it Chengdu Institute of Computer Applications, Chinese
Academy of Sciences, Chengdu,\\ Sichuan 610041, PR China \quad
E-mail: luyang@casit.ac.cn}}\ec

\vskip 1 mm

\noindent{\small {\small\bf Abstract} \ \ This note shows the equivalence of two projection
 operators which both can be used in cylindrical algebraic decomposition (CAD) . One is known as
 Brown's Projection (C. W. Brown (2001)); the other was proposed by Lu Yang in his earlier
 work (L.Yang and S.~H. Xia (2000)) that is sketched as follows: given a polynomial $f$ in
  $x_1,\,x_2,\,\cdots$, by $f_1$ denote the resultant of $f$ and its partial derivative with
   respect to $x_1$ (removing the multiple factors), by $f_2$ denote the resultant of $f_1$ and
    its partial derivative with respect to $x_2$, (removing the multiple factors), $\cdots$,
    repeat this procedure successively until the last resultant becomes a univariate polynomial.
  Making use of an identity, the equivalence of these two projection operators is evident.

\vspace{1mm}\baselineskip 12pt

\no{\small\bf Keywords} \ \
cylindrical algebraic decomposition (CAD), projection operator, resultant.

\no{\small\bf MR(2000) Subject Classification}\  12Y05, 13P15, 14P10, 68W30\ {\rm }}

\sec{1\quad Introduction}

Cylindrical algebraic decomposition (CAD) is a key constructive tool
in real algebraic geometry. Given such a decomposition in $\mathbb{R}^n$
it is easy to obtain a solution of a given semi-algebraic system. The CAD-algorithm was established by
collins [1] in 1975 as the basis of his quantifier elimination method
in real closed fields and has been improved over the years (see [2],[3],[4],[5],[6],[7]).

CAD construction proceeds in two phases, projection and lifting. The
projection phase computes a set of polynomials called {\it
projection factor set}\/ which consists of the irreducible factors
of given polynomials. The focus of this note is projection phase,
while lifting phase is not described. The readers are referred to
(Collins, 1988 [1]; Collins and Hong,1991 [4]) for a detailed
description of the lifting phase.

\sec{2\quad An identity involving resultants, discriminants and  leading coefficients}

We require a number of well-known definitions which can be found in some books ([8], [9]).

Let $f$ and $g$ be two non-zero polynomials of degree $n$ and $m$ in $\mathbb{R}[x]$.
\begin{eqnarray*}
f=a_n x^n+\cdots+a_0,\\
g=b_m x^m+\cdots+b_0.
\end{eqnarray*}
We define the Sylvester matrix associated to $f$ and $g$ and the
resultant of $f$ and $g$\/ as follows.

\indent{\bf Definition 1}  The Sylvester matrix of $f$ and $g$, denoted by
$\mathbf{S}(f,g)$, is the matrix
$$
\mathbf{S}(f,g)=
\begin{pmatrix}
a_n & \cdots & \cdots & \cdots & \cdots & a_0 & 0     & \cdots & 0 \\
0& \ddots &  &        &        &         &\ddots & \ddots &\vdots\\
\vdots & \ddots & \ddots &        &        &        &   & \ddots &0\\
0 & \cdots & 0 &   a_n     & \cdots & \cdots & \cdots & \cdots & a_0 & \\
b_m & \cdots & \cdots & \cdots & b_0 & 0 & \cdots   & \cdots & 0 \\
0& \ddots &  &        &        &     \ddots    &  \ddots    &  & \vdots\\
\vdots & \ddots & \ddots &        &        &        &  \ddots    & \ddots  & \vdots \\
\vdots &  & \ddots &   \ddots     &        &        &     & \ddots  & 0 \\
0 & \cdots & \cdots &   0    &     b_m   &   \cdots     &  \cdots   &  \cdots  & b_0\\
\end{pmatrix}.
$$
It has $m+n$ columns and $m+n$ rows.
The $\mathbf{resultant}$ of $f$ and $g$, denoted by $\mathrm{Res}(f,g,x)$, is the determinant of $\mathbf{S}(f,g)$.

The other formula on $\mathrm{Res}(f,g,x)$ is as follows.
If $a_n\neq 0$ and $b_m\neq 0$ then
\begin{equation}
\mathrm{Res}(f,g,x)=a_n^n b_m^m\prod_{i=1}^n \prod_{j=1}^m (x_i-y_j)=(-1)^{mn}\mathrm{Res}(g,f,x),
\end{equation}
where $x_1,\ldots, x_n$ are roots of $f$ and $y_1,\cdots y_m$ are roots of $g$ in $\mathbb{C}$.

\indent{\bf Definition 2} Let $f\in \mathbb{R}[x]$ be a polynomial of degree $n$,
$$f=a_n x^n+a_{n-1} x^{n-1}+\cdots+a_0,$$
and let $x_1,\ldots,x_n$ be the roots of $f$ in $\mathbb{C}$ (repeated according to their multiplicities).
The  $\mathbf{discriminant}$ of $f$, $\mathrm{Dis}(f)$, is defined by
\begin{equation}
\mathrm{Dis}(f,x)=(-1)^{\frac{n(n-1)}{2}}a_n^{2n-2}\prod_{n\geq i> j \geq 1}(x_i-x_j)^2.
\end{equation}

The following proposition holds clearly from (1) and Definition 2.

By $f'$ denote the derivative of $f$, here
$$f'=na_nx^{n-1}+\cdots+a_1.$$
Then we have the following formula
\begin{equation}
\mathrm{Res}(f,f',x)=a_n \mathrm{Dis}(f,x)=\mathrm{Lc}(f,x)\cdot \mathrm{Dis}(f,x),
\end{equation}
where $\mathrm{Lc}(f,x)$ is the leading coefficient.

Using the equations (1) and (2) the following equation  may be easily proved,

\begin{equation}
\mathrm{Res}(fg,(fg)',x)=\mathrm{Res}(f,f',x)\cdot\mathrm{Res}(g,g',x)\cdot
\mathrm{Res}(f,g,x)\cdot \mathrm{Res}(g,f,x).
\end{equation}
The equation (4) is also equivalent to the classical formula ( I. M. Gelfand,
M. M. Kapranov, A. V. Zelevinsky (1994), P405,(1.32) [8] )
\begin{equation*}
\mathrm{Res}(f,g,x)^2=(-1)^{mn}\frac{\mathrm{Dis}(fg,x)}{\mathrm{Dis}(f,x)\mathrm{Dis}(g,x)}.
\end{equation*}

\indent{\bf Example 1.} Let $f=ax^2+bx+c$ and $g=a_1x^2+c_1$ in $\mathbb{R}[x]$.
Then the equation (4) is the following.

\begin{align*}
& \left |
\begin{pmatrix}
aa_1 & ba_1 & (ac_1+ca_1) &bc_1& cc_1 & 0 & 0  \\
0& aa_1 & ba_1 & ac_1+ca_1 &bc_1& cc_1 & 0   \\
0& 0& aa_1 & ba_1 & ac_1+ca_1 &bc_1& cc_1  \\
4aa_1 & 3a_1b & (2ac_1+2a_1c)& bc_1 & 0 & 0& 0\\
0 &4aa_1 & 3a_1b & (2ac_1+2a_1c)& bc_1 & 0 & 0\\
0&0 &4aa_1 & 3a_1b & (2ac_1+2a_1c)& bc_1 & 0 \\
0&0&0 &4aa_1 & 3a_1b & (2ac_1+2a_1c)& bc_1  \\
\end{pmatrix}
\right | \\
& =
\left |
\begin{pmatrix}
a & b & c  \\
2a &b &0 \\
0 & 2a & b\\
\end{pmatrix}
\right |
\cdot
\left |
\begin{pmatrix}
a_1 & 0 & c_1  \\
2a_1 &0 &0 \\
0 & 2a_1 & 0\\
\end{pmatrix}
\right |
\cdot
\left |
\begin{pmatrix}
a & b & c &0  \\
0 & a & b & c  \\
a_1 &0 & c_1 &0\\
0 & a_1 & 0 & c_1\\
\end{pmatrix}
\right |
\cdot
\left |
\begin{pmatrix}
a_1 &0 & c_1 &0\\
0 & a_1 & 0 & c_1\\
a & b & c &0  \\
0 & a & b & c  \\
\end{pmatrix}
\right |
\end{align*}

\indent{\bf Remark 1.} Note that the equation of Example 1 is an identity relative to
$a, b, c, a_1, c_1$. In other words, the equation (4) still holds when we replace
the ring $\mathbb{R}[x]$ with the ring $\mathbb{R}[x_1,\cdots, x_n]$ and the
 variable $x$ with $x_i$. The book
 Algebra (M. Artin (1991), pages 456-457. [12]) has a very brilliant discussion
  how to give a strict proof of this kind of identities.\\

We present a key Lemma now .

\indent{\bf Lemma 1}  Let $s\geq 2, k \geq 2$ and let $f, f_1, \cdots, f_s $
be squarefree polynomials of positive degree in $\mathbb{R}[x_1,\ldots, x_k]$. Then
the following two identities hold,
 \begin{equation}
\mathrm{Res}(f,\frac{\partial (f)}{\partial x_1},x_1)
=\mathrm{Lc}(f,x_1)\cdot
 \mathrm{Dis}(f,x_1).
\end{equation}
 \begin{equation}
\mathrm{Res}(\prod_{i=1}^s f_i,\frac{\partial (\prod_{i=1}^s f_i)}{\partial x_1},x_1)
=\prod_{i=1}^s\mathrm{Lc}(f_i,x_1)\cdot \prod_{i=1}^s
 \mathrm{Dis}(f_i,x_1)\cdot \prod_{i\neq j\leq s}\mathrm{Res}(f_i,f_j,x_1).
\end{equation}
Where $\mathrm{Lc}$, $\mathrm{Dis}$ and $\mathrm{Res}$ denote the leading coefficients,
 discriminants and resultants respectively.\\

\indent{\bf Proof}\quad It is clear that the identity (5) comes from (3).

 The following is the proof of (6) by induction on the number of polynomials $s$.

Base case: $s=2$.

From (4),\ (5) and Remark 1, we have
 \begin{align*}
&\mathrm{Res}(f_1 f_2,\frac{\partial (f_1 f_2)}{\partial x_1},x_1)\\
&=\mathrm{Res}(f_1,\frac{\partial (f_1)}{\partial x_1},x_1)\cdot
\mathrm{Res}(f_2,\frac{\partial (f_1)}{\partial x_1},x_1)\cdot
 \mathrm{Res}(f_1,f_2,x_1)\cdot \mathrm{Res}(f_2,f_1,x_1)\\
&=\mathrm{Lc}(f_1,x_1)\cdot \mathrm{Lc}(f_2,x_1)\cdot \mathrm{Dis}(f_1,x_1)\cdot \mathrm{Dis}(f_2,x_1)
\cdot \mathrm{Res}(f_1,f_2,x_1)\cdot \mathrm{Res}(f_2,f_1,x_1).
\end{align*}

Induction step: suppose $s=t>2$, the following holds,
 \begin{align}
\mathrm{Res}(\prod_{i=1}^t f_i,\frac{\partial (\prod_{i=1}^t f_i)}{\partial x_1},x_1)
=\prod_{i=1}^t\mathrm{Lc}(f_i,x_1)\cdot \prod_{i=1}^t
 \mathrm{Dis}(f_i,x_1)\cdot \prod_{i\neq j \leq t}\mathrm{Res}(f_i,f_j,x_1).
\end{align}

From (5) we have the identity
\begin{align}
\mathrm{Lc}(\prod_{i=1}^tf_i,x_1)\cdot\mathrm{Dis}(\prod_{i=1}^tf_i,x_1)
=\mathrm{Res}(\prod_{i=1}^t f_i,\frac{\partial (\prod_{i=1}^t f_i)}{\partial x_1},x_1).
\end{align}

From the base case s=2 we have
 \begin{align*}
&\mathrm{Res}(\prod_{i=1}^{t+1} f_i,\frac{\partial (\prod_{i=1}^{t+1} f_i)}{\partial x_1},x_1)\\
&=\mathrm{Lc}(\prod_{i=1}^tf_i,x_1)\cdot \mathrm{Lc}(f_{t+1},x_1)\cdot \mathrm{Dis}
(\prod_{i=1}^tf_i,x_1)\cdot \mathrm{Dis}(f_{t+1},x_1)\cdot \mathrm{Res}(\prod_{i=1}^tf_i,f_{t+1},x_1)\\
&\ \ \cdot \mathrm{Res}(f_{t+1},\prod_{i=1}^tf_i,x_1)\\
&=\mathrm{Res}(\prod_{i=1}^t f_i,\frac{\partial (\prod_{i=1}^t f_i)}{\partial x_1},x_1)\cdot \mathrm{Lc}(f_{t+1},x_1)
\cdot\prod_{i=1}^{t+1} \mathrm{Dis}(f_i,x_1)\cdot \mathrm{Res}(\prod_{i=1}^tf_i,f_{t+1},x_1) \\
&\ \ \cdot \mathrm{Res}(f_{t+1},\prod_{i=1}^tf_i,x_1).
\end{align*}

We need yet the multiplicativity of resultant ([8]).
\begin{align}
&\mathrm{Res}(\prod_{i=1}^tf_i,f_{t+1},x_1)=\prod_{i=1}^t\mathrm{Res}(f_i,f_{t+1},x_1).\\
&\mathrm{Res}(f_{t+1},\prod_{i=1}^tf_i,x_1)=\prod_{i=1}^t\mathrm{Res}(f_{t+1},f_i,x_1).
\end{align}

Using the inductive hypothesis (7) and the equations (9) , (10), we obtain at once that
\begin{align*}
\mathrm{Res}(\prod_{i=1}^{t+1} f_i,\frac{\partial (\prod_{i=1}^{t+1} f_i)}{\partial x_1},x_1)
=\prod_{i=1}^{t+1}\mathrm{Lc}(f_i,x_1)\cdot \prod_{i=1}^{t+1}
 \mathrm{Dis}(f_i,x_1)\cdot \prod_{i\neq j \leq t+1}\mathrm{Res}(f_i,f_j,x_1).
\end{align*}
The proof is completed.

\sec{3\quad The equivalence of two projection for CAD  }

The first projection  $\mathbf{Proj}$ comes from C.W. Brown (2001) [6].

\indent{\bf Definition 3 } (C. W. Brown (2001))   Let $A$ be a squarefree basis in $\mathbb{Z}[x_1,\cdots,x_k]$,
where $k\geq 2$. Definite the projection Proj($A, x_j$) of $A$ to be the union
of the set of all leading coefficients of elements of $A$ in variable $x_j$, the set of all discriminants
of elements $f$ of $A$  in variable $x_j$ , and the set of all resultants of pairs $f$ ,
 $g$ of distinct elements of $A$ in variable $x_j$ .\\

 $P\longleftarrow \mathbf{Projection\ 1}(A)$ \quad ($\mathrm{Proj}$ , \quad C.W. Brown (2001))

 Input: $A\subseteq \mathbb{R}[x_1,\ldots,x_k]$

 Output: $P$, the projection factor set of a sign-invariant CAD for $A$.\\

\qquad (1) $P_0$:= IrreducibleFactorsOf$(A)$, $P:=P_0$.

\qquad (2) for $j$ from $1$  to $k-1$ do

\qquad  \qquad  $P_{j}$:= IrreducibleFactorsOf$(\mathrm{Proj}(P_{j-1},x_j))$

\qquad  \qquad  $P:=P\bigcup P_{j}$.

\qquad (3) Return $P$.\\

The other projection  $\mathbf{ResP}$ comes from L. Yang {\it et al}. (2000, 2001) [11],[12].

\indent{\bf Definition 4} (L. Yang {\it et al}. (2000)) Let $A$ be a squarefree basis
 in $\mathbb{Z}[x_1,\cdots,x_k]$,
where $k\geq 2$. Definite the projection ResP($A,x_j$) of $A$ to be the squarefree polynomial,
which is the product of irreducible factors of the resultant
$$\mathrm{Res}(\prod_{ \alpha\in A} \alpha,\ \frac{\partial(\prod_{\alpha\in A} \alpha)}{\partial x_j},x_j).$$\\

$P\longleftarrow \mathbf{Projection\ 2}(A)$\quad ($\mathrm{ResP}$)

 Input: $A\subseteq \mathbb{R}[x_1,\ldots,x_k]$

 Output: $P$, the projection factor set of a sign-invariant CAD for A.\\

\qquad (1) $P_0$:= IrreducibleFactorsOf$(A)$, $P:=P_0$.

\qquad (2) for $j$ from 1 to $k-1$ do

\qquad  \qquad $P_{j}$:= IrreducibleFactorsOf$(\mathrm{ResP}(P_{j-1},x_j))$

\qquad  \qquad  $P:=P\bigcup P_{j}$.

\qquad (3) Return $P$.\\

According to Lemma 1, it is clear that the projection Proj and ResP
generate the same projection factor set. So they are equivalent for CAD.

\sec{4\quad ResP for one polynomial }

The projection  ResP is firstly applied to solving the global optimization problems
for algebraic functions by L. Yang {\it et al}. (2000, 2001)[11],[12]. Recently,
 J. J. Han, L. Y. Dai, B. C. Xia (2014) [7]
improve the method by adding GCD computation. In the same paper, they have proved
that improved HP (ResP) still guarantees obtaining at least one sample point from every
 connected component of the highest dimension.

Using  the projection ResP to only one polynomial $f\in\mathbb{R}[x_1,\cdots,x_n]$,
 we can usually obtain a triangular system, the real zeros of which contain all of the
 real critical points of $f$. Namely the real zeros of the system
 $$
 \left \{
 \begin{array}{llll}
f=0,\\
\frac{\partial(f)}{\partial x_1}=0,\\
\cdots\\
\frac{\partial(f)}{\partial x_n}=0.
\end{array}
\right.
 $$
must satisfy the following triangular system
$$
 \left \{
 \begin{array}{llll}
f=0,\\
\mathrm{ResP}(f,x_1)=0,\\
\cdots\\
\mathrm{ResP}(\cdots \mathrm{ResP}(\mathrm{ResP}(f,x_1),x_{2})\cdots,x_{n-1})=0.
\end{array}
\right.
 $$

For the polynomial $f-T$\ (where $f\in \mathbb{R}[x_1,\ldots,x_n]$), the equation
\begin{equation}
\mathrm{ResP}(\cdots \mathrm{ResP}(\mathrm{ResP}(f-T,x_1),x_{2})\cdots,x_{n})=0
\end{equation}
is of only a variable T. If the optimal value of function $f$ exists in $\mathbb{R}^n$,
 say $f_{min}$, then $f_{min}$
 has to satisfy equation (11).

\no \vskip0.2in
\no {\bf References}
\vskip0.1in

\footnotesize

  \REF{[1]} G. E. Collins. Quantifier elimination for real closed fields by cylindrical
 algebraic decomposition. In {\it Second GI Conference Automata Theory and Formal Languages,
  Kaiserslauten}, Volume 33 of Lecture Notes Comp. Sci., pages 134-183. Springer, 1975.

  \REF{[2]} S. McCallum. An improved projection operation for cylindrical algebraic
   decomposition of three-dimensional space. {\it J. Symb. Comput.} 5(1): 141-161, 1988.

  \REF{[3]} H. Hong. A Improvement of the projection operator in cylindrical algebraic
   decomposition. In {\it proceedings ISSAC'90}, 261-264, 1990.

  \REF{[4]} G. E. Collins, H.Hong. Partial cylindrical algebraic decomposition for
   quantifier elimination. {\it J. Symb. Comput.}, 12(3): 299-328, 1991.

  \REF{[5]} S. McCallum. An improved projection operation for cylindrical algebraic decomposition.
  In {\it Quantifier Elimination and Cylindrical Algebraic Decomposition}, 242-268, Springer,1998.

  \REF{[6]} C. W. Brown. Improved projection for cylindrical algebraic decomposition.
{\it J. Symb. Comput.} 32(5): 447-465, 2001.

  \REF{[7]} J. J. Han, L. Y .Dai, B. C. Xia. Constructing fewer open cells by GCD computation in
  CAD projection. In {\it Proc. ISSAC'14}, 240-247, 2014.

  \REF{[8]} I. M. Gelfand, M. M. Kapranov, A. V. Zelevinsky. Discriminants, Resultants and
  Multidimensional Determinants. Boston: Birkh$\ddot{a}$user, 1994.

  \REF{[9]} L. Yang, B. Xia, Automated Proving and Discovering on
Inequalities (in Chinese). Beijing: Science Press, 2008.

 \REF{[10]} M. Artin. Algebra. Englewood Cliffs: Prentice Hall, 1991.

  \REF{[11]}  L. Yang. S. H. Xia. An inequality-proving program applied to global optimization. In
{\it Proceedings of ATCM 2000}, W-C. Yang {\it et al} (eds.), ATCM, Inc., Blacksburg, 40-51, 2000.

(Also see http://epatcm.any2any.us/EP/EP2000/ATCMP208/fullpaper.pdf)

  \REF{[12]} L. Yang. A symbolic algorithm for optimization and finiteness principle. In
 Mathematics and Mathematics Mechanization, D. D. Lin {\it et al }(eds). Jinan:
  Shandong Educational Publishing House, 2001.

(Also see http://book.lh.chaoxing.com/ebook/detail.jhtml?id=10924129\&page=210)

\end{document}